\documentstyle[12pt]{article}

\def\fun#1#2{\lower3.6pt\vbox{\baselineskip0pt\lineskip.9pt

\ialign{$\mathsurround=0pt#1\hfil##\hfil$\crcr#2\crcr\sim\crcr}}}

\newcommand{\be}{\begin{equation}}

\newcommand{\ee}{\end{equation}}

\newcommand{\bd}{\begin{displaymath}}

\newcommand{\ed}{\end{displaymath}}

\newcommand{\ba}{\begin{array}}

\newcommand{\ea}{\end{array}}

\newcommand{\bt}{\begin{tabular}}

\newcommand{\et}{\end{tabular}}

\begin{document}

\vspace{1cm}
\begin{flushright}
PSU/TH/203
\end{flushright}

\vspace{1cm}

\begin{center}

{\Large\bf

HEAVY QUARK LIMIT IN THE LIGHT FRONT QUARK MODEL}\\[1.2cm]

N.B. Demchuk \\

Penn State University, 104 Davey Lab, University Park PA 16802, U.S.A.\\

\end{center}

\vspace{1cm}

\begin{abstract}
An explicit relativistic light-front quark model is presented which gives
the momentum transfer dependent form factors of weak hadronic currents
among heavy pseudoscalar and vector mesons in
the whole accessible kinematical region $ 0\leq q^2 \leq q^2_{max} $.
It is shown that in the limit of infinite masses of active quarks 
these form factors can be expressed in terms of universal Isgur- Wise function.
The explisit expression for this function is obtained. 
It is shown that
neglect of pair creation from the vacuum in calculations of
form factors does not violate Luke's theorem.
\end{abstract}

\newpage

\section{Introduction}

One of the most interesting subjects in the investigation of the Standard
Model is the study of CP violation effects and the determination of
electroweak theory fundamental parameters. In the coming decade 
the main efforts in this direction will be applied in the heavy
quark sector. A fundamental problem for theory is to extract data at quark
level from experiments that involve hadrons.

Since in the infinite quark mass limit the spins of the heavy and light 
degrees of freedom decouple QCD experiences great simplifications.
A new $SU(2N_h)$ spin-flavour symmetry that QCD reveals for $N_h$ heavy-quark
species, \cite{VS}, \cite{IW}, \cite{Hussain} appears. 
This symmetry, which is not manifest
in the original QCD lagrangian, becomes explicit in an effective field theory,
the so-called heavy-quark effective theory \cite{Georgi}, \cite{KT91}, 
\cite{MRR}.
In this limit Isgur and Wise have derived many
simple and appealing relations and normalization conditions for various 
hadronic matrix elements. For semileptonic transition between two heavy mesons 
they have got that all the assosiated hadronic form factors can be expressed 
in terms of single universal function $\xi(y)$ 
(where $y=u_1\cdot u_2$ and $u_{1\mu},~u_{2\mu}$ are the four-velocities of 
the heavy meson
before, after the transition respectively), the Isgur-Wise function. 
In order to make direct connection between heavy hadron and the corresponding
quark amplitudes we need knowledge of $\xi(y)$. It only depends
on the transfer of four-velosities of the heavy mesons and 
is normalized at zero recoil. This single form factor incorporates all of
the effects of the interaction between the heavy and light degrees of freedom.
Its theoretical understanding is therefore of great interest. Since 
this function is sensetive to the effects of QCD at large distances, 
it cannot  
be calculated in pertubation theory. 

The statement that the dynamics of processes
involving transitions among (infinetly) heavy quarks depends only on 
the velocity transfer has some interesting consequences. 
Although the Isgur-Wise function itself is not determined from symmetry,
restrictive relations between various hadronic form factors arise.
In this work we use them to test the consistency of light front (LF)
constituent quark model which is presented 
in a series of papers \cite{DGNS96},\cite{CCH97},\cite{DKNO97}.

The main assumption of this approach was neglect of pair creation from
the vacuum in calculation of the weak decay form factors. It resulted 
in slight dependence of the form factors upon the choice of the reference
frame. Though the heavy quark limit in the relativistic light-front
quark model has already been studied \cite{CCHZL}, it is a good consistent
check to show that this dependence vanishes in this limit. The investigation
of $ \frac{1}{m_Q} $ expansion of matrix elements of hadronic currents
(where $ m_Q $ is the mass of heavy quark) can shed light on how small
pair creation contribution is for heavy-to-heavy transitions. The new
result in this work is that above mentioned procedure does not violate 
Luke's theorem \cite{Luke}, which deals with the first order corrections
to the infinite quark mass limit. 

The plan of the paper is as follows.
In Sections 2-4 we describe our approach for the calculation of the weak
meson form factors and show that in the limit of infinite masses
of the active quarks our results obey the pattern of heavy quark symmetry. 
In Section 5 we show that the expressions for the form factors obtained
in \cite {DKNO97} satisfy Luke's theorem. 
In Section 6 the numerical results for the Isgur-Wise function are presented
and compared with the results of other 
approaches. Section 7 contains a brief summary.

\section{Kinematics}

Herebelow, we denote by $ P_1 $,$ P_2 $ and $ M_1 $, $ M_2 $ the 4-momentum
and masses of the parent and daughter mesons, respectively. The meson states
are denoted as $|P>$ for a pseudoscalar state and $|P,\varepsilon>$ for a
vector state, where $\varepsilon$ is the
polarization vector, satisfying $\varepsilon\cdot P=0$.
The 4-momentum transfer $q$ is given by $ q=P_1-P_2 $ and
the momentum fraction $r$ is defined as

\be
r=\frac{P_2^+}{P_1^+}=1-\frac{q^+}{P^+_1}.
\ee
We work in the rest frame of the parent meson.
The { \it 3}-momentum
$ \vec P_2 $ of the final meson is in the plane 1-3, so that
$ {\bf q_{\bot}} \not= 0 $. We denote the angle
between $ \vec P_2 $ and the 3-axis by $ \alpha $. Then, it can be easily
verified that

\be
\label{2}
q^2=(1-r)(M_1^2-\frac{M_2^2}{r})-\frac{q^2_{\bot}}{r},
\ee

\be
q^2_{\bot }=M_2^2(y^2-1)sin^2\alpha,
\ee
where the "velocity transfer" $ y $ is defined as $ y=u_1u_2 $
with $ u_1 $ and $ u_2 $ being the 4-velocities of the initial and
final mesons. The relation between $ y $ and  $ q^2 $ is given by
\be
\label{3}
 y =\frac{M^2_1+M^2_2-q^2}{2M_1M_2}.
\ee
The momentum fraction $ r $ is invariant under the boosts
along the $ 3-$ axis and the rotations around this axis but depends
explicitely on the recoil direction. Solving Eq.(\ref{2}) for $r$ one obtains
\be
r(q^2,\alpha )=\zeta (y+\sqrt{y^2-1}cos \alpha),
\ee
where $ \zeta=\frac{M_2}{M_1} $.
Note that at the point of zero recoil $r$ does not depend on $\alpha $,
$ r(q^2_{max})=\zeta $.

From Lorenz invariance one finds the form factor decomposition
of matrix elements of the vector and axial currents.
We define the form factors of the
$ P_1(Q_1 \bar q) \to P_2(Q_2 \bar q) $  transitions between the ground state
$S$-wave mesons in the usual way.  The amplitude
$<P_2|V_{\mu}|P_1>=<P_2|\bar Q_2\gamma_{\mu}Q_1|P_1>$ can be
 expressed in terms of two form factors

\be
\label{4}
<P_2|V_{\mu}|P_1>=\left(P_{\mu}-\frac{M^2_1-M^2_2}{q^2}
q_{\mu}\right)F_1(q^2)+\frac{M^2_1-M^2_2}{q^2}q_{\mu}F_0(q^2),
\ee
where  $ P=P_1+P_2 $. There is one form factor for the amplitude
$ <P_2,\varepsilon,|V_{\mu}|P_1> $

\be
\label{5}
<P_2,\varepsilon|V_{\mu}|P_1>=\frac{2i}{M_1+M_2}
\varepsilon_{\mu\nu\alpha\beta}\varepsilon^{*\nu}P^{\alpha}_1
P^{\beta}_2V(q^2),
\ee
and three independent form factors for the amplitude

$ <P_2,\varepsilon|A_{\mu}|P_1>=
<P_2,\varepsilon|\bar Q_2\gamma_{\mu}\gamma_5Q_1|P_1> $

\begin{eqnarray}
\label{6}
<P_2,\varepsilon|A_{\mu}|P_1> & = &
\left((M_1+M_2)\varepsilon^{*\mu}
A_1(q^2)-\frac{\varepsilon^*q}{M_1+M_2}(P_1+P_2)_{\mu}A_2(q^2)-
2M_2\frac{\varepsilon^*q}{q^2}q_{\mu}A_3(q^2)\right) \nonumber \\
+2M_2\frac{\varepsilon^*q}{q^2}q_{\mu}A_0(q^2),
\end{eqnarray}
where $ A_0(0) = A_3(0) $ and $ A_3(q^2) $ is given by the
linear combination
\be
A_3(q^2)=\frac{M_1+M_2}{2M_2}A_1(q^2)-\frac{M_1-M_2}{2M_2}A_2(q^2).
\ee

In the case of heavy--to--heavy transitions, in the limit in which
the active quarks have infinite mass,
all the form factors are given in terms of a single function $\xi(y)$, the
Isgur--Wise form factor. In the realistic case of finite quark masses these
relations are modified: each form factor depends separately on the dynamics
of the process.

The relations between the form factors arising in this limit read

\be
F_1=V=A_0=A_2=\frac{M_1+M_2}{2\sqrt{M_1M_2}} \xi(y),
\ee

\be
F_0=A_1=\frac{1+y}{M_1+M_2} \sqrt{M_1M_2} \xi(y).
\ee

\section{The matrix elements of the vector and axial currents}

In ref. \cite{DKNO97} it was shown that one can determine all of the form
factors by taking matrix elements of the good components of the weak
currents. These components may be written in the form \footnote{The spectator
quark carries the fraction $x$ of the plus component of the meson momentum,
while the heavy quark carries the fraction $1-x$. In what follows it is assumed
that the variables $r$, $q^2$, and $q^2_{\bot}$ are related
by equation (2) }
\be
\label{8}
J_V(q^2,r)=<P_2|V^+|P_1>=\sqrt{M_1M_2}\int\limits^{r(\alpha)}_0
\frac{dx}{2x}\int \frac{d^2 k_{\bot}}{m^2}\phi_2(x', k'^2_{\bot})
\phi_1(x, k^2_{\bot})
\cdot I_V,
\ee
\be
\label{9}
J_{V,+1}(q^2,q_{\bot})=
<P_2,\varepsilon(+1)|V^+|P_1>=\sqrt{M_1M_2}\int\limits^{r(\alpha)}_0
\frac{dx}{2x}\int \frac{d^2k_{\bot}}{m^2}\phi_2(x', k'^2_{\bot})
\phi_1(x, k^2_{\bot})
\cdot I_{V,+1},
\ee
\be
\label{10}
J_{A,0}(q^2,r)=
<P_2,\varepsilon(0)|A^+|P_1>=\sqrt{M_1M_2}\int\limits^{r(\alpha)}_0
\frac{dx}{2x}\int \frac{d^2k_{\bot}}{m^2}\phi_2(x', k'^2_{\bot})
\phi_1(x, k^2_{\bot})
\cdot {\it I}_{A,0},
\ee
\be
\label{11}
J_{A,+1}(q^2,q_{\bot})=<P_2,\varepsilon(+1)|A^+|P_1>=\sqrt{M_1M_2}
\int\limits^{r(\alpha)}_0
\frac{dx}{2x}\int \frac{d^2k_{\bot}}{m^2}\phi_2(x', k'^2_{\bot})
\phi_1(x, k^2_{\bot})
\cdot {\it I}_{A,+1},
\ee

where $I_V$, $I_{V,+1}$, $I_{A,\rho}$, $(\rho=0,+1)$ are contributions of 
the Dirac currents and quark
spin structures:

\be
\label{aa}
I_V=\frac{\mu_1\mu_2}{2mM_1M_2} 
Tr[R^+_{00}(x',k'_{\bot}\bar \lambda, \lambda_2)
\bar u(\bar p_2,\lambda_2)
\gamma^+u(p_1,\lambda_1)R_{00}(x,k_{\bot},\lambda_1,\bar \lambda)],
\ee

\begin{eqnarray}
\label{bb}
I_{V,+1}=\frac{\mu_1\mu_2}{2mM_1M_2}
Tr[R^+_{1+1}(x',k'_{\bot},\bar \lambda, \lambda_2)
\bar u(\bar p_2,\lambda_2)\gamma^+u(p_1,
\lambda_1)R_{00}(x,k_{\bot}\lambda_1,\bar \lambda)],
\end{eqnarray}

\begin{eqnarray}
\label{cc}
I_{A,\rho}=\frac{\mu_1 \mu_2}{2mM_1M_2}
Tr[R^+_{1,\rho}(x',k'_{\bot},\bar \lambda, \lambda_2)
\bar u(\bar p_2,\lambda_2)\gamma^+\gamma_5u(p_1,\lambda_1)
R_{00}(x,k_{\bot},\lambda_1,\bar \lambda)],
\end{eqnarray}
with
\be
\mu_1 = [M^2_{10}-(m_1-m)^2]^{1/2},~~~~
\mu_2= [M^2_{20}-(m_2-m)^2]^{1/2},
\ee
and
 \be
    M_{10}^2  \equiv  M_{10}^2(x,  k^2_{\bot}) = \frac{m^2 +
     k^2_{\bot}}{x} + \frac{m^2_1 +  k^2_{\bot}}{1-x},
 \ee
 \be
    M_{20}^2  \equiv  M_{20}^2(x',  k'^2_{\bot}) = \frac{m^2 +
     k'^2_{\bot}}{x'} + \frac{m_2^2 +
 k'^2_{\bot}}{1-x'}. \\ \label{19}
 \ee
In these equations $m_1$ and $m_2$ are the masses of active quarks, $m$ is
the mass of the quark--spectator, $x'=\frac{x}{r}$ and
${\bf k'_{\bot}}={\bf k_{\bot}}+x'{\bf q_{\bot}}$. In our kinematics
\be
k_1'= k_1-x'\sqrt{y^2-1} M_2 sin\alpha, ~~~k_2'=k_2.
\ee
The spin wave functions $R_{J,J_3}$ can be found in ref. \cite{Jaus}.
The wave function $\phi(x,k^2_{\bot})$ can be related to wave function
$\chi(x,k_{\bot}^2)$ from \cite{DKNO97} by a simple formula
$$
\phi(x,k^2_{\bot})=2\sqrt{M_im}m\chi(x,k_{\bot}^2).
$$
It has been shown in \cite{DGNS96} that phenomenological wave function
$\chi(x,k_{\bot}^2)$ can be written in terms of equal-time wave function 
$w(k^2) $ normalized according to      
 \be
 \label{18}
    \int_0^{\infty} dk ~ k^2 ~ w^2(k^2) = 1,
\ee
where the fraction $x$ is replaced by the relative longitudinal momentum
$ k^{(1)}_3 $ of two quarks in the parent meson defined as
 \be
    k^{(1)}_3 = \left( x - {1 \over 2} \right) M_{10} + \frac{m_1^2 -
    m^2}{2M_{10}},
 \ee
and the fraction $x'$ is replaced by the relative 
longitudinal momentum $ k^{(2)}_3 $ of two quarks in daughter meson
 \be
    k^{(2)}_3 = \left( x' - {1 \over 2} \right) M_{20} + \frac{m_2^2 -
    m^2}{2M_{20}}.
 \ee
Explicitly, one has \cite{DGNS96}
 \be
 \label{21}
    \phi_i(x,  k^2_{\bot}) = \frac{\sqrt{M_im}m}{1 - x} \frac{\sqrt{M_{i0} [1 -
    (m_i^2 - m^2)^2 / M^4_{i0}]}} {\sqrt{M^2_{i0} - (m_i - m)^2}} ~
    \frac{w_i(k^2)}{\sqrt{4\pi}},
 \ee
with $k^2 \equiv k_{\bot}^2 + k_3^2$. 

Noting that $\bar u(p_2,\lambda_2)\gamma^+ u(p_1,\lambda_1)
=\sqrt{4p_1^+p_2^+}\delta_{\lambda_2 \lambda_1}$,\\
 $\bar u(p_2,\lambda_2)\gamma^+\gamma_5 u(p_1,\lambda_1)
=\sqrt{4p_1^+p_2^+}\varphi^+_{\lambda_2} \sigma_3 \varphi_{\lambda_1}$,
where $ \varphi_{\lambda} $ are the Pauli spinors
we obtain
\be
I_V
=\frac{1}{x'mM_2}[{\cal A}_1{\cal A}_2+{\bf k_{\bot}k'_{\bot}}],
\ee
\be
I_{V,+1}
=-\frac{1}{\sqrt2mx'M_2}
[k'_1{\cal A}_1-k_1{\cal A}_2+\frac{2k^2_2(k'_1-k_1)}{M_{20}+m+m_2}],
\ee
\be
I_{A,+1}
=\frac{1}{\sqrt2mx'M_2}[(2x'-1)k'_1{\cal A}_1+k_1{\cal A}_2+2k'_1
\frac{{\cal A}_1{\cal B}+{\bf k_{\bot} k'_{\bot}}}{M_{20}+m+m_2}],
\ee
\be
I_{A,0}
=\frac{1}{mx'M_2}[{\cal A}_1{\cal A}_2+(1-2x'){\bf k_{\bot}k'_{\bot}}+
\frac{2{\cal A}_1k'^2_{\bot}-2{\cal B}({\bf k_{\bot}k'_{\bot}})}
{M_{20}+m+m_2}],
\ee
with

\be
 {\cal A}_1=xm_1+(1-x)m, ~~ {\cal A}_2=x'm_2+(1-x')m,
\ee
and
\be
 {\cal B}=(1-x')m-x'm_2.
\ee

To extract a leading term in $(\frac{1}{m_Q})$ expansions of good components 
of the currents it should be noted that the effective wave function 
$ \phi_i(x,k_{\bot}^2) $ has the maximum at $x=\frac{m}{m_i}$ and its width
$\approx \frac{m}{m_i}$ \cite{DK}. Thus for heavy-to-heavy 
transitions quantities $\frac{m}{M_1}$, $\frac{m}{M_2}$, 
$\frac{M_1-m_1}{M_1}$, $\frac{M_2-m_2}{M_2}$, $x$, $x'$, 
should be treated as small parameters. 
In the leading term of this expansion we get
\be
\label{26}
I_V^{(0)}
=\frac{1}{x'mM_2}[(m+xM_1)(m+x'M_2)+{\bf k_{\bot}k'_{\bot}}]=
\left(1+\frac{(v,u_1+u_2)}{1+y}\right)(u_1+u_2)^+,
\ee
$$
I_{V,+1}^{(0)}
=\frac{1}{\sqrt2mx'M_2}
(k'_1(xM_1+m)-k_1(x'M_2+m))=
$$
\be
\label{27}
-i\epsilon^{+\alpha \beta \gamma}(u_{1\alpha}
u_{2\beta} \varepsilon^*_{\gamma}(+1)+u_{1\alpha}
v_{\beta} \varepsilon^*_{\gamma}(+1)+v_{\alpha}
u_{2\beta} \varepsilon^*_{\gamma}(+1)),
\ee
$$
I_{A,+1}^{(0)}
=-\frac{1}{\sqrt2mx'M_2}
(k'_1(xM_1+m)-k_1(x'M_2+m))=
$$
\be
\label{28}
-u_2^+(v\varepsilon^*(+1)+u_1\varepsilon^*(+1))
-v^+(u_1\varepsilon^*(+1))+u_1^+(v\varepsilon^*(+1)),
\ee
$$
I_{A,0}^{(0)}
=\frac{1}{x'mM_2}[(m+xM_1)(m+x'M_2)+{\bf k_{\bot}k'_{\bot}}]=
$$
\be
\label{29}
\left(1+\frac{(v,u_1+u_2)}{1+y}\right)(u_1+u_2)^+,
\ee
where index (0) denotes the leading term of the expansion,
$u_1$,$u_2$ are four-velocities of parent and daughter mesons 
respectevily and $v$ is four-velocity of spectator-quark
\be
v=(v^-,v^+,v_{\bot})=(\frac{m^2+k_{\bot}^2}{xM_1m}, \frac{xM_1}{m},
\frac{k_{\bot}}{m}).
\ee

$\varepsilon(\lambda), (\lambda=\pm 1,0)$ are polarization vectors:

\be
\varepsilon(0)=\frac{1}{M_2}\left(\frac{-M_2^2+P^2_{\bot}}{P^+},P^+,P_{\bot}
\right),~~ \varepsilon(\pm 1)=\left(
\frac{2}{P^+}{\bf P_{\bot}\epsilon_{\bot}}(\pm 1),0,\epsilon_{\bot}(\pm 1)
\right),
\ee

where

$$
\epsilon_{\bot}(\pm 1)=\mp \frac{1}{\sqrt2}(1,\pm i).
$$

Noting that 

\be
(k^2_i)^{(0)}=m^2[(vu_i)^2-1],
\ee

\be
  \left(\frac{\sqrt{M_i}}{1 - x} \frac{\sqrt{M_{i0} [1 -
   (m_i^2 - m^2)^2 / M^4_{i0}]}} {\sqrt{M^2_{i0} - (m_i - m)^2}} \right)^{(0)}=
   \frac{\sqrt{2vu_i}}{\sqrt{1+vu_i}},
 \ee
we may write a leading term in $(\frac{1}{m_Q})$ expansion of the effective 
wave function $\phi_i(x,k_{\bot}^2)$
\be
\label{32}
\left(\phi(x,k_{\bot}^2)\right)^{(0)}=\phi^{(0)}(vu_i)=
m^{3/2}\frac{\sqrt{2vu_i}}{\sqrt{1+vu_i}}\frac{w^{(0)}(m[(vu_i)^2-1])}
{\sqrt{4\pi}},
\ee
where 
\be
w^{(0)}(m^2[(vu_i)^2-1])=\lim_{m_i \rightarrow \infty} w_i(k_i^2).
\ee
In the rest frame of meson when the heavy quark mass is infinite the inner 
momentum ${\bf \vec k }$ is equal to {\it 3}-momentum of light quark and 
normalization condition for $w^{(0)}(m^2\vec v^2)$ is
\be
m^3\int \frac{d^3\vec v}{4\pi}[w^{(0)}(m^2\vec v^2)]^2=1.
\ee
It can be easily verified that $\frac{dxd^2k_{\bot}}{2xm^2}=\frac{d^3v}{2v^0}$.
Using eqs. (\ref{26} -\ref{29}), (\ref{32}) and an obvious fact that 
\be
\int \frac{d^3v}{2v^0}\phi^{(0)}(vu_1)\phi^{(0)}(vu_2)v_{\alpha}=
\int \frac{d^3v}{2v^0}\phi^{(0)}(vu_1)\phi^{(0)}(vu_2)
\frac{(v,u_1+u_2)}{2(1+y)}(u_1+u_2)_{\alpha},
\ee
we may write for the good components of the currents the following 
expressions
\be
J^{(0)}_V(q^2,r)=\sqrt{M_1M_2}\xi (y) (1+\frac{rM_1}{M_2}),
\ee
\be
J^{(0)}_{V,+1}(q^2,q_{\bot})=\sqrt{\frac{M_1}{M_2}}\xi (y) q_{\bot},
\ee
\be
J^{(0)}_{A,0}(q^2,r)=\sqrt{M_1M_2}\xi (y) (1+\frac{rM_1}{M_2}),
\ee
\be
J^{(0)}_{A,+1}(q^2,q_{\bot})=\frac{1}{\sqrt2}
\sqrt{\frac{M_1}{M_2}}\xi (y) q_{\bot},
\ee
where $\xi (y) $ is universal Isgur-Wise function 
\be
\label{37.a}
\xi(y)=\int \frac{d^3v}{2v^0}\phi^{(0)}(vu_1)\phi^{(0)}(vu_2)
\left(1+\frac{(v,u_1+u_2)}{1+y}\right).
\ee
It has the same structure as that one obtained
from the analysis of the Feynman triangle diagram assuming simple exponential
parametrization for the vertex functions $\phi^{(0)}(vu_i)$ \cite{DK}.
This function is normalized at the point of zero recoil 
\be
\xi(1)=m^3\int \frac{d^3\vec v}{4\pi}[w^{(0)}(m^2\vec v^2)]^2=1.
\ee

In case of the spinless quarks (i.e. assuming $R_{00}=1$
and $\bar u(\bar p_2) \gamma^+ u(p_1)=p^+_1+p^+_2$ in eq. (\ref{aa}))
it can be easily shown that universal form factor $\xi_{w.s.}(y)$ 
takes the form
\be
\label{37.b}
\xi_{w.s.}(y)=\int \frac{d^3v}{2v^0}\phi^{(0)}(vu_1)\phi^{(0)}(vu_2)
\sqrt{1+vu_1} \sqrt{1+vu_2}.
\ee

\section{The definition of the vector and axial form factors in the infinite
quark mass limit}

Recall \cite{DGNS96} that the time-like LF result for the "good" 
component of the weak
vector current $ J^+=J^0+J^3 $ coincides with
the contribution of the spectator pole of the Feynman triangle diagram,
which corresponds to the valence quark approximation, while
the remaining part of the Feynman diagram, the so-cold Z graph, can not
be expressed directly in terms of a valence quark wave function.
The sum of both contributions does not depend, of course, on the choice of 
the frame but each contribution is frame-dependent. Therefore the time-like 
LF result for the form factors generally depends on
the recoil direction of the daughter meson relative to the 3-axis e.g. on
the choice of angles $ \alpha $ specifying a reference frame.
Fortunately Z graph does not give contribution to the leading term of
$(\frac{1}{m_Q})$
expansion of plus components of weak currents and thus 
in this limit the form factors do not depend on this choice. To illustrate 
this point we reexamine an approach of ref. \cite{DGNS96}, which was used
to calculate the form factors $F_1(Q^2)$ for PS--PS transitions.We first reexamine an approach of ref. \cite{DGNS96} to calculate 
the form factors $ F_1(q^2) $ for the $ PS-PS $ transitions. 
Equation (\ref{4}) for the plus component of the vector current 
yields only one constraint
for the two formfactors $ F_1(q^2) $ and $ F_0(q^2) $. In order to 
revert Eq. (\ref{4}) the matrix element of the current was calculated in 
\cite{DGNS96} in two reference frames having the 3-axis 
parallel and anti-parallel to the {\it 3}-momentum of the daughter meson.
This corresponds to the choice $\alpha_1=0,\alpha_2=\pi$ where the angles 
$ \alpha_i $ have been defined in Section 2. But we can use two other frames.
Specifying these frames by the two arbitrary 
angles $ \alpha_1 $ and $ \alpha_2 $ we write Eq. (35) of ref. \cite{DGNS96}
as
\be
\label{39}
F_1(q^2)=\frac{(1-r(\alpha_2))J_V(q^2,r(\alpha_1))-
(1-r(\alpha_1))J_V(q^2,r(\alpha_2))}
{2M_1(r(\alpha_1)-r(\alpha_2))}.
\ee
In the leading term of $\frac{1}{m_Q}$ expansion we get for the form factor
$F_1(q^2)$

$$
F_1(q^2)=\frac{(1-r(\alpha_2))J_V^{(0)}(q^2,r(\alpha_1))-
(1-r(\alpha_1))J_V^{(0)}(q^2,r(\alpha_2))}
{2M_1(r(\alpha_1)-r(\alpha_2))}=
$$
\be
\label{40}
=\frac{M_1+M_2}{2\sqrt{M_1M_2}}\xi (y).
\ee
So in this limit the slight dependence of the form factor upon the $\alpha$
choice  vanishes. Using the equations (13), (15), (48), (50) and (52)
of ref. \cite{DKNO97} we may apply the same procedure for form factors
$ V, A_1, A_2, A_0 $.
\be 
\label{41}
V^{(0)}(q^2)=\frac{1}{\sqrt2}(1+\zeta)[\frac{\partial}{\partial q_{\bot}}
J^{(0)}_{V,+1}(q^2,q_{\bot})]=\frac{M_1+M_2}{2\sqrt{M_1M_2}}\xi(y),
\ee
\be
\label{42}
A^{(0)}_0(q^2)=\frac{1}{2M_1r}J^{(0)}_{A,0}(q^2,r)-
\frac{(1-r)\zeta}{\sqrt2r}A^{(0)}_+(q^2,r)-
\frac{(\zeta^2-r^2)(1-r)}{2\sqrt2\zeta}
\frac{\partial}{\partial r}A_+^{(0)}(q^2,r)= 
\frac{M_1+M_2}{2\sqrt{M_1M_2}}\xi(y),
\ee
\be
\label{43}
A^{(0)}_1(q^2)=\frac{1}{M_1+M_2}\frac{\zeta}{r}[J^{(0)}_{A,0}(q^2,r)
\frac{M_2}{\sqrt2\zeta^2}(r^2-\zeta^2)A^{(0)}_+(q^2,r)]= 
\frac{1+y}{M_1+M_2}\sqrt{M_1M_2}\xi(y),
\ee
\be
\label{44}
A^{(0)}_2(q^2)=\frac{M_1+M_2}{\sqrt2 M_1}(A^{(0)}_+(q^2,r)+
r(1-r)\frac{\partial}{\partial r}A^{(0)}_+(q^2,r))
=\frac{M_1+M_2}{2\sqrt{M_1M_2}}\xi(y),
\ee
where
$A^{(0)}_+(q^2,r)=\frac{\partial}{\partial q_{\bot}}
J^{(0)}_{A,+1}(q^2,q_{\bot})$ and $\zeta = \frac{M_1}{M_2}$.
Thus we have shown that all hadronic form factors for semileptonic 
transitions between two heavy mesons can be expressed in terms of a single
universal function $\xi (y)$.

\section{Luke's theorem} 

In what follows we show that LF approach results for the weak decay 
form factors \cite{DKNO97} satisfy Luke's theorem. The only new parameter
that enters the calculation is the difference between the meson mass and
the mass of the heavy quark 
$$
\bar \Lambda =M_B-m_b=M_D-m_c=M_{D^*}-m_c,
$$
where $M_{D^*}=M_D$ up to terms of order $\frac{1}{m_c}$ \cite{NR}.
As noted by Luke \cite{Luke} the hadronic matrix elements in 
equations (\ref{4} - \ref{6} )
are not affected by $\frac{1}{m_Q}$ corrections
at the zero-recoil normalization point $u_1=u_2$. Inasmuch as the kinematical
sructures for $J_{V,+1}$, $J_{A,+1}$ vanish at this point, Luke's theorem
implies 
\be
\label{45}
J^{(1)}_V((M_1-M_2)^2,\zeta)=J^{(1)}_{A,0}((M_1-M_2)^2,\zeta)=0,
\ee
where index (1) denotes leading corrections to the infinite quark mass limit.
Expanding equations (\ref{8}, \ref{10}) as a power series in $(\frac{1}{m_Q})$
at the point of zero recoil $u_1=u_2=u$ we get 
$$
J^{(1)}_V((M_1-M_2)^2,\zeta)=
$$
\be
\label{46}
=\sqrt{M_1M_2}\int_0^{\zeta}\int 
\frac{dxd^2k_{\bot}}{2xm^2}\left((\phi^{(0)}(vu))^2I^{(1)}_V+
\phi^{(0)}(vu)\phi^{(1)}_1(x,k_{\bot}^2)I^{(0)}_V+
\phi^{(0)}(vu)\phi^{(1)}_2(x',k_{\bot}^2)I^{(0)}_V \right),
\ee 
$$
J^{(1)}_{A,0}((M_1-M_2)^2,\zeta)=
$$
\be
\label{47}
=\sqrt{M_1M_2}\int_0^{\zeta}\int 
\frac{dxd^2k_{\bot}}{2xm^2}\left((\phi^{(0)}(vu))^2I^{(1)}_{A,0}+
\phi^{(0)}(vu)\phi^{(1)}_1(x,k_{\bot}^2)I^{(0)}_{A,0}+
\phi^{(0)}(vu)\phi^{(1)}_2(x',k_{\bot}^2)I^{(0)}_{A,0} \right),
\ee
where
\be
\label{48}
I^{(1)}_V=I^{(1)}_{A,0}=-(\frac{1}{M_1}+\frac{1}{M_2})
\frac{m+\bar \Lambda}{m}(xM_1+m).
\ee
To treat the leading correction to $\phi^{(0)}(vu)$ one should consider
the expansion of normalization condition for the effective wave function 
$\phi_i(x,k_{\bot}^2)$ equations (\ref{18},\ref{21})
\be 
\label{49}
\int^1_0\int \frac{dxd^2k_{\bot}}{4xm^2}\phi^2_i(x,k^2_{\bot})I_i=1,
\ee
where $I_i=\frac{1}{xmM_i}[A^2_i+k^2_{\bot}]$, $i=1,2$.
Expanding this condition we get
\be
\label{50}
1=\int^1_0\int\frac{dxd^2k_{\bot}}{4xm^2}(\phi^{(0)}(vu))^2I^{(0)}+
\int^1_0\int\frac{dxd^2k_{\bot}}{4xm^2}
\left(2\phi^{(0)}(vu)\phi^{(1)}_i(x,k^2_{\bot})I^{(0)}+
(\phi^{(0)}(vu))^2I^{(1)}_i\right)+...,
\ee
where
$$
I^{(0)}=\frac{1}{xmM_1}((xM_1+m)^2+k^2_{\bot})=2(1+vu),
$$
$$
I^{(1)}_i=-2\frac{1}{M_i}\frac{m+\bar \Lambda}{m}(xM_i+m),
$$
and the omitted terms in (\ref{50}) will give next power corrections.
Noting that 
\be
\int^1_0\int\frac{dxd^2k_{\bot}}{4xm^2}(\phi^{(0)}(vu))^2I^{(0)}=\xi (1)=1,
\ee
we get from equation (\ref{50}) the equation for $\phi^{(1)}_i(x,k^2_{\bot})$
\be
\int^1_0\int\frac{dxd^2k_{\bot}}{4xm^2}
\left(2\phi^{(0)}(vu)\phi^{(1)}_i(x,k^2_{\bot})I^{(0)}+
(\phi^{(0)}(vu))^2I^{(1)}_i\right)=0.
\ee
Substituting it into equations (\ref{46}, \ref{47}) we obtain
\be
J^{(1)}_V((M_1-M_2)^2,\zeta)=\sqrt{M_1M_2}\int_0^{\zeta}\int 
\frac{dxd^2k_{\bot}}{2xm^2}(\phi^{(0)}(vu))^2(I^{(1)}_V-
\frac{1}{2}I^{(1)}_1-\frac{1}{2}I^{(1)}_2)=0,
\ee
\be
J^{(1)}_{A,0}((M_1-M_2)^2,\zeta)=\sqrt{M_1M_2}\int_0^{\zeta}\int 
\frac{dxd^2k_{\bot}}{2xm^2}(\phi^{(0)}(vu))^2(I^{(1)}_V-
\frac{1}{2}I^{(1)}_1-\frac{1}{2}I^{(1)}_2)=0,
\ee
which is in agreement with equation (\ref{45}).

\section{Results}

The calculation of the Isgur-Wise function (eq. (\ref{37.a})) has been 
performed using equation (\ref{32}) for meson wave function 
$\phi^{(0)}(vu_i)$. As for the radial wave function $w^{(0)}$ appearing
in equation (\ref{32}) we used two different ans\"{a}tze. The authors
of ref. \cite{MYOPR} found, for three relativistic models \cite{GI85},
\cite{CNP}, \cite{VD}, the wave functions are rather close to each other 
and that they are rather well reproduced by an exponential form in $r$:
$$
w(r)=2a^{-3/2}exp(-r/a),
$$

\be
w(k^2)=\sqrt{\frac{32}{\pi}}a^{3/2}\frac{1}{(1+a^2k^2)^2},
\ee
with $a^{-1}=0.75~ GeV$.
The light quark mass $m$ is not well determined in the above relativistic 
spectroscopic models. But at the same time $\xi(y)$ and $\rho^2$ are not very 
sensitive to $m$. This is due to the fact that 
light quark is ultrarelativistic. In our calculations we used $m=0.3~ GeV$.
The second ansatz for the radial wave function is the Gaussian of the 
Isgur-Scora-Grinstein-Wise (ISGW) model, 
 the values of the parameters (the mass of the light constituent
quark and the harmonic oscillator (HO) length) are taken from 
\cite{ISGW89}.

The $y$ behaviour of the Isgur-Wise function is shown in Fig. 1. The solid
and dashed lines are the results of our LF calculation with the first and
the second ans\"{a}tze for the radial wave function respectively. 
For comparision, the result obtained
from the analysis of the Feynman triangle diagram assuming simple exponential
parametrization for the vertex functions $\phi^{(0)}(vu_i)$ \cite{DK}
is shown by dotted line.

For small, non-zero, recoil it is conventional to write 
$$
\xi(y)=1-\rho^2(y-1)+O((y-1)^2),
$$
where $ \rho^2 $ is the slope of the Isgur Wise function at zero recoil.
Using the first ansatz we get $ \rho^2=1.0 $, which reproduces
the result of 
the relativistic quark models of form factors 
a l\'{a} Bakamjian-Thomas \cite{MYOPR}  and is in agreement 
with predictions
of QCD fundamental methods.
QCD sum rules have been used to calculate the slope parameter of the Isgur-Wise
function; the results obtained by various authors are $\rho^2=0.84 \pm 0.02$
(Bagan {\it et al.} \cite{Bagan}), 
$0.7 \pm 0.1 $ (Neubert \cite{Neubert}), $0.70 \pm 0.25$ (Blok and Shifman
\cite{Blok}) and $1.00 \pm 0.02$ (Narison \cite{Narison}).
UKQCD lattice calculation \cite{Bowler} yields 0.9 as central value,
admittedly with very large error bars:
\be
\rho^2=0.9^{+0.2+0.4}_{-0.3-0.2}.
\ee
For the second ans\'{a}tz we obtained considerably higher value $\rho^2=1.4$,
however it is in excellent agreement with calculation by Close and Wambach
 \cite{CW94}.

Two aspects \cite{MYOPR} account of the difference between our two ans\"{a}tze
results. The main aspect is that the first ans\"{a}tz corresponds to the 
relativistic spectroscopic models while the second to nonrelativistic one.
In ref. \cite{MYOPR} it was found that wave functions at rest can be strongly
different in a relativistic spectroscopic model from those of nonrelativistic,
although the spectrum is similar, whence lowering of $\rho^2$.

Another, more technical, but important aspect is that the approximation of
using a variational Gaussian to approximate the wave function fails even at
low recoil, e.g. in the calculation of $\rho^2$, especially in the context
of relativistic spectroscopic equations. Calculating exactly the wave function 
results in appreciably lower estimate of $\rho^2$.

Performing the calculations with and without the effects of the Melosh
composition (eqs. (\ref{37.a},\ref{37.b})), 
it turns out that the effects of the Melosh
rotations increase $\rho^2$ by $ 30\% $ and $ 20\% $ for the first
and the second ans\"{a}tze respectively; last result agrees with 
the conclusion of ref. \cite{CW94} obtained in the zero-binding approximation.

\section{Conclusions}

In relativistic quark model formulated on the light front we have examined
$\frac{1}{m_Q}$ expansion of form factors for semileptonic transitions
between mesons. We verified that in leading term of this expansion all
associated hadronic form factors can be expressed in terms of universal 
Isgur-Wise function. Explisit expression has been given for it.
Using the fact that form factors are related to overlap integrals of hadronic
wave functions we have shown that this function is normalized at the point
of zero recoil. We have shown that the relativistic light front quark model
of form factors, combained with a relativistic spectroscopic model
to calculate the needed wave functions, reproduces the result
for the slope of the universal form factor ($\rho^2 \approx 1$) 
of relativistic quark models a l\'{a} Bakamjian-Thomas.
This result is in agreement with QCD fundamental methods.
 In next-to-leading term we obtained that our results obeyed
Luke's theorem. Before closing, it should be reminded that in our analysis the
contribution of the pair creation from the vacuum has been neglected.
Therefore we can conclude that it does not contribute to the leading
term of $\frac{1}{m_Q}$ expansion of the form factors 
and at least at the point of zero recoil
to next-to-leading one.

\section*{Acknowledgement}

We are grateful to D. Melikhov and I.M. Narodetskii for many enlightening 
discussions.

\newpage

\section*{\bf Figure Caption}
\begin{description}

\item[Fig.1]  The Isgur-Wise function $\xi (y) $.
The relation between the kinematical
variables $y$ and $q^2$ is given by equation (\ref{3}).
 The solid
and dashed lines are the results of our LF calculation with the first and
the second ans\"{a}tze for the radial wave function respectively. 
 The dotted line is  the result obtained
from the analysis of the Feynman triangle diagram assuming simple exponential
parametrization for the vertex functions $\phi^{(0)}(vu_i)$ \cite{DK}.

\end{description}

\newpage
\begin{figure}[htb]
\vskip 9.0in\relax\noindent\hskip -.5in\relax{\includegraphics{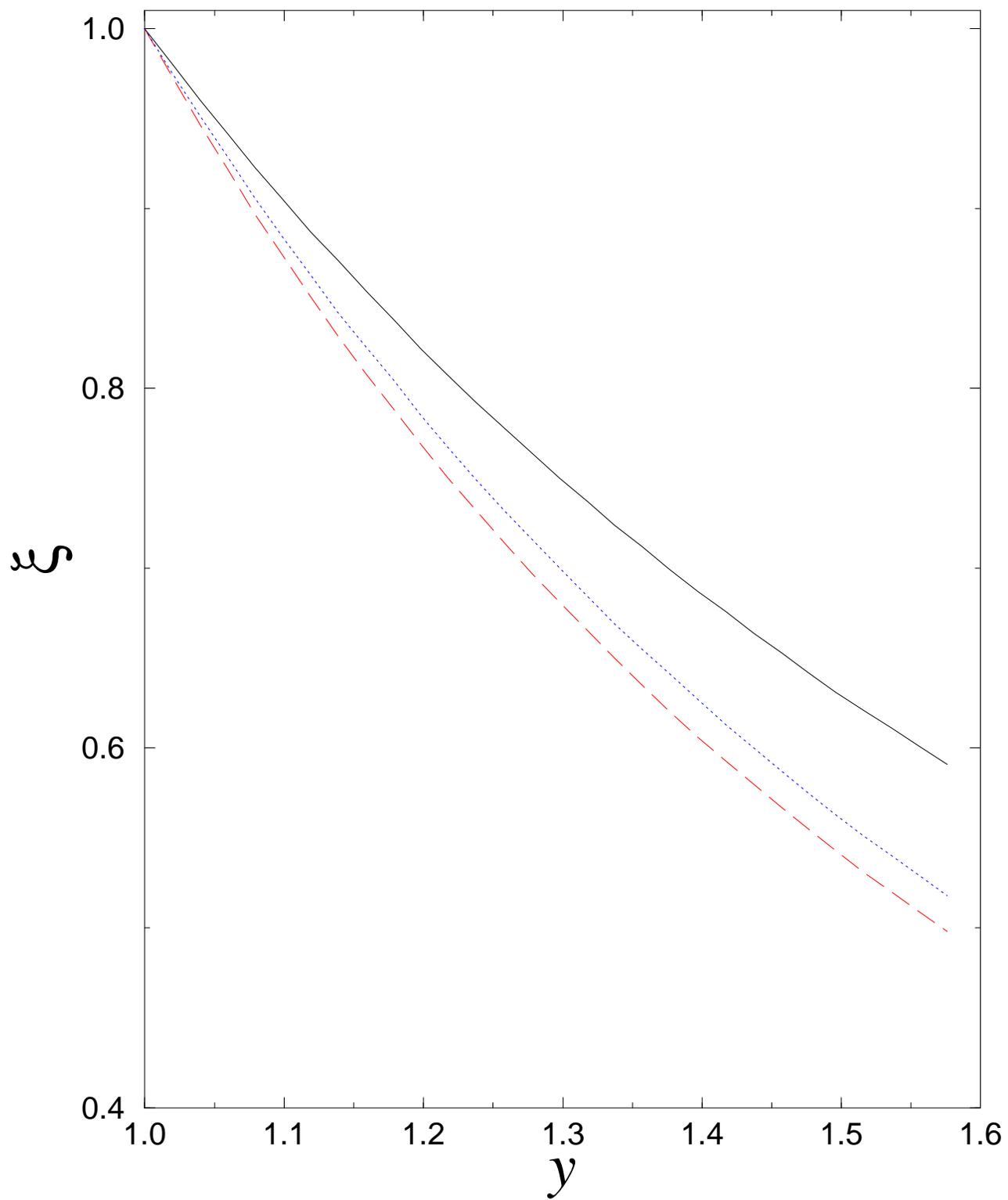}}
\caption{The Isgur-Wise function $\xi(y)$.}
 \end{figure}
\end{document}